\documentclass[sn-mathphys-num]{sn-jnl}

\usepackage{graphicx}%
\usepackage{multirow}%
\usepackage{amsmath,amssymb,amsfonts}%
\usepackage{amsthm}%
\usepackage{mathrsfs}%
\usepackage[title]{appendix}%
\usepackage{xcolor}%
\usepackage{textcomp}%
\usepackage{manyfoot}%
\usepackage{booktabs}%
\usepackage{algorithm}%
\usepackage{algorithmicx}%
\usepackage{algpseudocode}%
\usepackage{listings}%
\usepackage{siunitx}
\usepackage{comment}
\usepackage{hyperref}
\usepackage{pdflscape}
\usepackage{makecell}
\usepackage{multirow}
\usepackage{csquotes}
\usepackage{censor}

\begin{document}

\title[Article Title]{Trust and Human Autonomy after Cobot Failures: Communication is Key for Industry 5.0}

\author*[1]{\fnm{Felix} \sur{Glawe}}\email{glawe@comm.rwth-aachen.de}

\author[2]{\fnm{Laura} \sur{Kremer}}\email{laura.kremer@ika.rwth-aachen.de}

\author[1]{\fnm{Luisa} \sur{Vervier}}\email{vervier@comm.rwth-aachen.de}

\author[1]{\fnm{Philipp} \sur{Brauner}}\email{brauner@comm.rwth-aachen.de}

\author[1]{\fnm{Martina} \sur{Ziefle}}\email{ziefle@comm.rwth-aachen.de}

\affil[1]{\orgdiv{Chair for Communication Science}, \orgname{RWTH Aachen University}, \orgaddress{\street{Campus-Boulevard 57}, \city{Aachen}, \postcode{52074}, \country{Germany}}}
\affil[2]{\orgdiv{Institute for Automotive Engineering}, \orgname{RWTH Aachen University}, \orgaddress{\street{Steinbachstraße 7}, \city{Aachen}, \postcode{52074}, \country{Germany}}}

\abstract{Collaborative robots (cobots) are a core technology of Industry 4.0. Industry 4.0 uses cyber-physical systems, IoT and smart automation to improve efficiency and data-driven decision-making. Cobots, as cyber-physical systems, enable the introduction of lightweight automation to smaller companies through their flexibility, low cost and ability to work alongside humans, while keeping humans and their skills in the loop. Industry 5.0, the evolution of Industry 4.0, places the worker at the centre of its principles: The physical and mental well-being of the worker is the main goal of new technology design, not just productivity, efficiency and safety standards. Within this concept, human trust in cobots and human autonomy are important. While trust is essential for effective and smooth interaction, the workers' perception of autonomy is key to intrinsic motivation and overall well-being. As failures are an inevitable part of technological systems, this study aims to answer the question of how system failures affect trust in cobots as well as human autonomy, and how they can be recovered afterwards. Therefore, a VR experiment (n = 39) was set up to investigate the influence of a cobot failure and its severity on human autonomy and trust in the cobot. Furthermore, the influence of transparent communication about the failure and next steps was investigated. The results show that both trust and autonomy suffer after cobot failures, with the severity of the failure having a stronger negative impact on trust, but not on autonomy. Both trust and autonomy can be partially restored by transparent communication.}


\maketitle

\section{Introduction}\label{Introduction}
The fourth industrial revolution, Industry 4.0, refers to the interconnection between machines and production systems enabling increased efficiency and flexibility \cite{lasi2014industry}. While the technologies driving Industry 4.0 are steadily being deployed in today's production landscape, the European Commission put forth the vision of Industry 5.0. The main idea of Industry 5.0 is to develop a resilient, sustainable, and human-centred industry, harnessing the advances made within Industry 4.0 but to an end that goes beyond the sole increase in efficiency \cite{i50eu, alves2023}. In production, the physical and mental well-being of workers shall be achieved by providing a motivating and safe environment that meets workers' needs, including factors such as autonomy, privacy and dignity. \cite{passalacqua2024human, alves2023, pusztahelyi2024improving}. Industry 5.0 can therefore be viewed as the challenge of balancing human well-being and automation and its metrics of success \cite{alves2023}. 

Key technologies of Industry 4.0 include cloud technologies, smart sensors, simulation, AI and advanced robotics, including collaborative robots (cobots) \cite{cimini2020, zheng2021}. Compared to traditional robots, cobots can operate in close proximity to a human partner and are designed to be highly flexible \cite{ISO_Cobot}. They are also designed to enhance the capabilities of human workers and not to replace them \cite{liu2024human}. Due to their high flexibility, they can easily be adapted to the operators' needs making them the ideal connecting element between Industry 4.0 and 5.0 \cite{alves2023}. Two measures are crucial to ensure the well-being of the worker and the effectiveness of the human-robot collaboration (HRC): appropriate trust in the collaboration and the satisfaction of the workers' need for autonomy \cite{esterwood2022literature, bennett2023}. Trust between a human and a robotic partner is necessary to ensure the adoption and appropriate use of the technology that enables the full potential of this relationship \cite{lotz_2019, panchetti_2023, zhang_2023}. This is particularly evident in industrial workplaces, where cobots are increasingly employed. In such contexts, trust becomes even more critical, as the misuse or non-use of these cobots can compromise the safety of workers and also undermine the profitability and efficiency of the technology \cite{kopp_2024, lee_2004, lewis_2018}. The need for autonomy, on the other hand, is a basic psychological need that must be satisfied to ensure the workers' well-being and to foster intrinsic motivation \cite{ryan2018}. However, designing the interaction between humans and robots (HRI) to preserve and foster human autonomy seems to be a non-trivial task. For example, Nikolova et al. 2022 reported a decrease in workers' perceived autonomy with increasing levels of industrial robotisation. Nevertheless, it is theorized that robots could increase human autonomy by replacing dull, repetitive or even dangerous tasks \cite{NIKOLOVA2024104987, turja2022basic}.\\
The question arises of how HRC must be designed to induce trust and fulfil the need for autonomy. One aspect that is expected to play a crucial role is the cobot's behaviour after a mistake. In human-human relationships, mistakes and unexpected behaviours can occur due to misunderstanding or miscommunication, and in the case of technology, because of faulty software, hardware, or misuse \cite{honig2018understanding}. However, these errors or failures can affect human trust in a cobot, and may also reduce human autonomy by compromising perceived system capabilities \cite{esterwood2022literature,friedman1996value}. To date, there is no consensus on the best strategies to restore trust after a failure and no study on the impact of failures on human autonomy. Therefore, it is important to investigate \textit{how failures in HRC affect human perception and how the cobot's behaviour after a failure can restore trust and autonomy to allow for future optimal collaboration and the workers' well-being.} This article addresses the research question through a virtual reality (VR) experiment that examines the influence of failures and their severity on trust perception and autonomy satisfaction. Furthermore, it investigates whether transparent communication can serve as a potential way to restore trust and autonomy after a failure.

In the first part of the article, the theoretical foundations of trust and human autonomy and previous findings on the relationship between trust, autonomy, failures and transparent communication will be discussed. In the second part, we'll derive the research question and hypotheses, and describe the method used to answer and test them. The results will be presented and discussed in the following two sections. Finally, we'll discuss the findings and limitations of the study.
\section{Related Work}\label{Related Work}
In the ensuing chapter, we introduce the concepts of trust in robots and human autonomy, together with the underlying theories. We then proceed to present recent findings related to these concepts, with particular reference to the possible factors contributing to the realisation of trust and autonomy in HRI and HRC. We subsequently explore the impact of failures in HRI and HRC on trust and autonomy, followed by an examination of the empirical findings and considerations on how both might be restored when violated. For those sections, we will derive the hypotheses respectively after each section. Given that HRC is a subtype of HRI, it is possible to apply the findings from HRI to HRC research. Therefore, we will present findings from both fields without repeatedly emphasising the need to confirm or refute findings for HRI in the more specific context of HRC.

\subsection{Trust}
Trust is a crucial element that must be thoroughly investigated and understood to ensure successful HRC \cite{dammers_usability_2022, tolmeijer_2020, kopp_2024}. A steadily increasing research interest in trust as a multidimensional component of HRC has resulted in a wide range of definitions and conceptions of trust. These vary depending on the context, including interpersonal trust between individuals, trust between individuals and organizations as well as trust between individuals and technology\cite{lee_2004, mcknight_2011}. Trust in HRI has thus far been examined using the same factors, such as reliability and competence, which are commonly discussed in the literature on human-automation interaction \cite{baker_2018, hancock_meta-analysis_2011, kessler_2017, malle_2021, tolmeijer_2020}. In the context of industrial use, cobots are consequently regarded as a component of industrial automation. In this regard, a distinction is drawn between automation and cobots, with the former being regarded as a more encompassing concept. It can be hypothesised that human-cobot trust and human-automation trust may possess analogous fundamental characteristics, though allowances must be made for the possibility of discrepancies \cite{hancock_meta-analysis_2011}. For the empirical study of trust in HRC, human trust in automation provides a basis on which to build \cite{baker_2018, hancock_meta-analysis_2011}.

Trust between humans and automated technology can be described as an attitude where it is assumed that an agent will assist in achieving an individual's goals in a situation marked by uncertainty and vulnerability \cite{lee_2004}. In HRI, trust is viewed as a relational concept that requires at least three elements. Two acting entities are needed: one acting entity to send information and another acting entity that is acted upon and receives the information \cite{hancock_2011}. A functioning communication channel between these two acting entities is the third necessary prerequisite \cite{baker_2018, hancock_2011}. Trust in HRI can be conceptualised as the human intention to rely on a robot, and it serves as a metric for the extent to which a human as a trustor believes in the actions and reliability of the robot as a trustee  \cite{zhu_2023}. The establishment, maintenance and calibration of human trust in a robot are crucial for successful collaboration  \cite{kok_2020, tolmeijer_2020}. 

In the context of establishing, maintaining and calibrating human-robot trust, it is implied that trust in HRI can be viewed from a time-dynamic perspective \cite{kopp_2024, tolmeijer_2020, kok_2020, pompe_2022, lewis_2018}. Trust can evolve from initial trust to knowledge-based trust in the course of the development of trust relationships \cite{kopp_2024, mcknight_2011, hoff2015trust}. Initial trust is based on the assessments of the trustors before they get to know the trustees \cite{mcknight_2011}. After gaining experience and familiarization with the trustees, trustors form a knowledge-based trust. Knowledge-based trust is shaped by internal and external variables such as self-confidence and task difficulty, as well as design features and system performance \cite{nayyar_2018, tolmeijer_2020}. Based on experienced interactions, the trustors know the trustees well enough to predict their behaviour in a given situation \cite{mcknight_2011}. Compared to initial trust, knowledge-based trust is more stable, as it is not subject to the weighing of motives and barriers \cite{mcknight_2011}. However, it has been demonstrated that individuals often exhibit an initial tendency to place trust in automated systems, even in the absence of prior experience \cite{nayyar_2018}.

The dynamic development of trust is assumed to be influenced by an interplay of characteristics of humans as trustors, robots as trustees and the communication channel \cite{hancock_2011}. An interaction takes place: the results of trust provide feedback to the trustors, potentially leading to adjustments or changes in their level of trust in the trustees \cite{hancock_2011}. This calibration of trust can influence trust in the system, the effectiveness of the HRC and thus the overall interaction \cite{hancock_2011}.

Various factors can influence the interaction described above. There has been growing interest in the assessment of factors that promote or hinder trust in HRC \cite{alarcon_2020,zhang_2023}. These factors are both extensive and complex. Essentially, the level of trust in HRC is determined by the dimensions of human characteristics (of the trustors), external environmental factors and robot characteristics (of the trustees) \cite{baker_2018, biermann_how_2021, hald_2021, hancock_2011, kessler_2017, lewis_2018, perkins_2022, tolmeijer_2020, zhu_2023}. 
Trust in robots is influenced by a variety of human characteristics, which include both ability-related and characteristic traits \cite{hancock_2011, hancock_meta-analysis_2011, lewis_2018}. Ability-related traits, such as attentiveness, expertise, competence, workload of the confidants, as well as their previous experience and situational awareness, are noteable examples of factors influencing trust \cite{hancock_2011, hancock_meta-analysis_2011, lewis_2018}. Characteristic traits can include demographic variables, personality traits, attitudes towards robots, comfort with robots, self-confidence and propensity to trust based on the trustors \cite{hancock_2011, hancock_meta-analysis_2011}. Ability-related and characteristic traits are seen as shaping the basic actions and psychological disposition of individuals towards robots and are therefore believed to also influence trust in robots \cite{hancock_2011, hancock_meta-analysis_2011}. 

External environmental factors can influence HRC, with factors such as teamwork, group affiliation, communication and culture being significant determinants \cite{hancock_meta-analysis_2011}. In previous studies, environmental factors have a moderate influence on trust, a phenomenon that can be attributed to the limited data available on external environmental factors \cite{hancock_meta-analysis_2011, hancock_2011, lewis_2018}.

Robot characteristics can be categorized into performance and moral characteristics in terms of the trustworthiness of the trustees. Performance trust refers to the robot's ability to perform a task and do so reliably, while moral trust refers to the robot's ability to perform the task and comply with social and moral norms honestly \cite{kopp_2024, lewis_2018, malle_2021, perkins_2022, zhang_2023}. In previous studies, both performance-related and attribute-based characteristics are defined in the trust-building process itself \cite{hald_2021, hancock_meta-analysis_2011}. Performance-related characteristics reflect the performance of the robot and refer, for example, to the robot's behaviour, reliability, demonstration capability, degree of automation and transparency. They play a decisive role in assessing the ability of a robot to perform a task or to perform it reliably \cite{hancock_2011, hancock_meta-analysis_2011}. Attribute-based properties include the spatial proximity, personality and adaptability of a robot, the robot type and anthropomorphic traits of the robot \cite{hancock_2011, hancock_meta-analysis_2011, nayyar_2018}. 

The distinction between performance and moral trust has parallels to the distinction between performance-based and attribute-based trust characteristics, in that performance-based characteristics of the robot can be related to performance-based trust expectations and attribute-based characteristics of the robot can be related to moral trust expectations. Earlier research has demonstrated that performance-related factors, particularly the predictability and reliability of a robot, are regarded as significant contributors to trust \cite{baker_2018, hancock_2011, malle_2021}.

The field of trust in HRI has been extensively studied and existing frameworks can be leveraged. However, rapid advancements in technology like HRC and the transformative shift towards Industry 5.0 as a new work paradigm highlight the remaining need for a more in-depth exploration of trust in these evolving contexts.

\subsection{Autonomy}
Autonomy is a concept that technological systems as well as humans are described to possess or lack. In its technical sense, autonomy signifies the extent to which a system can function autonomously, that is, with minimal reliance on external, human intervention. \cite{etzioni2016ai}, human autonomy however goes beyond this definition. Since Emanuel Kant laid his important groundwork human autonomy has been understood as a basic ethical principle \cite{schneewind1998invention, formosa2021robot}. Kant described autonomy as the freedom to act and choose without being dominated by others or one's own inclinations \cite{guyer2003kant}. While this may appear to suggest that autonomy is only attainable in the complete absence of dependence, it rather hinges on the alignment of the external input with one's values. If one merely obeys commands or requests, autonomy is absent but if the input is valuable and provides guidance to the recipient he or she can still be autonomous, even if it results in a degree of dependence \cite{ryan2018}. Besides human autonomy as a key moral and ethical principle, Ryan and Deci established the need for autonomy, in addition to the needs for relatedness and competence, as a basic psychological need. Satisfaction of those needs is key to intrinsic motivation, human flourishing and overall well-being \cite{ryan2018}. Within the theoretical framework of self-determination theory (SDT), a macro theory of human motivation, they describe the need for autonomy as \enquote{people’s universal urge to be causal agents, to experience volition, to act in accord with their integrated sense of self (i.e., with their interests and values), and to endorse their actions at the highest level of reflective capacity} (Deci and Vansteenkiste, 2004, p.5) \cite{deci2004self}. By establishing human autonomy as a psychological concept, they paved the way for empirical examination and measurement of autonomy. Complementing the important theoretical reflections with a psychological and empirical measured perception of autonomy can highlight problem zones where certain aspects are thought to enhance autonomy but are, in reality, not perceived as meaningful inputs \cite{spector1986perceived}. Therefore, the question of how satisfaction of those basic psychological needs can be achieved is examined in multiple, contexts including education, work environments and HRI \cite{jackson2016effect,nie2015importance, henkemans2017design}. Human autonomy is regarded as a foundational concept within human-computer interaction (HCI). Within the context of HCI, autonomy is frequently used interchangeably with the concept of agency. While these concepts are not entirely synonymous, they are intricately intertwined \cite{bennett2023}. In this article, the definition provided by Deci and Vansteenkiste will be the primary focus, however, we will draw on work related to both concepts, autonomy and agency, to provide a comprehensive perspective \cite{deci2004self}.  

As previously outlined, the fulfilment of the need for autonomy is associated with enhanced overall well-being, characterised by an augmentation in intrinsic motivation. Within the work context, other desirable outcomes are observable like job satisfaction, diminished staff turnover, heightened commitment among employees, and reduced exhaustion \cite{nie2015importance, van2010capturing, ahuja2007road}. Peters et al. established the METUX framework where technology use is theorized to impact multiple layers of experience in our life through satisfaction or thwarting of the basic psychological needs, from overall societal well-being down to the experience when interacting with the technologies interface \cite{peters2018designing}. The framework highlights the importance of careful technology design because technology can impact multiple facets of our lives. HRI as part of HCI is subject to the same considerations and multiple studies hypothesize how robots could satisfy our basic psychological needs. Robots have the potential to substitute for dangerous or monotonous tasks, thereby enabling more meaningful tasks or providing more valuable ends for example by assisting elderly people, extending their capabilities and therefore enhancing their autonomy \cite{NIKOLOVA2024104987, formosa2021robot}. However, research findings, particularly in the context of production, suggest that the integration of robots may potentially result in a reduction of human autonomy \cite{NIKOLOVA2024104987, turja2022basic, dammers_usability_2022}. Studies measuring human autonomy satisfaction in HRI confirmed the positive effects postulated by SDT. Task engagement, trust in the robot and well-being increased with higher autonomy satisfaction when interacting with robots \cite{van2020using, lee2023role, turja2022basic}. 

However, there is still a gap in the answer to the question of how robots and their interaction must be designed to retain and foster human autonomy. In a study conducted by Yu et al., the influence of path visualization by a delivery robot on a human's sense of agency was investigated. The robot could either visualize its own intended path for the pedestrian or the path that it predicted the human to take. Predicting the human's path led to a lower sense of agency because the participants may have felt unconsciously controlled by the robot. Conversely, the visualisation of the robot's trajectory led to an augmented sense of agency \cite{yu2023your}.
An additional noteworthy finding is reported by Kaur et al.: They led a survey with 530 blue collar workers on the effect of working with cobots. The study revealed that cobots, that only provided assistance when required, preserved autonomy, while it was lowest for cobots that never or always assisted \cite{kaur2023studying}. The personalisation of a robot has also been identified as a means to satisfy a user's need for autonomy. Henkemans et al. designed a robot playing an education game with children that provided task choices, and rationales for the learning task, and gave relevant feedback while addressing the children by their names. In comparison to the control group that played with a robot lacking these features, the autonomy satisfaction was higher \cite{henkemans2017design}. 

While this review shows that there are indeed ways to design HRI such that it supports the need for autonomy, research in this field is just starting to emerge.

\subsection{Failures in HRI}

Failures occur in diverse forms, and can be defined as \enquote{a degraded state of ability which causes the behavior or service being performed by the system to deviate from the ideal, normal, or correct functionality}(Honig \& Oron-Gilad, 2018, p.2)\cite{honig2018understanding}. Examples include inaccurate decoding of commands, failure to recognize presented objects or incorrect route planning \cite{zhang_2023}.

Four different types of failures caused by the users or the system can be distinguished in terms of their impact on trust: Design, System, Expectation and User failures \cite{tolmeijer_2020}. Design failures are failures that are not ideal or intuitive for the HRI due to the design in terms of the behaviour, appearance, and dialogue of a robotic system. System failures include unanticipated system behaviour of the hardware or software. Expectation failures refer to system behaviour that deviates from the user's expectations. User failures refer to user interaction with the system that is not intended and results in failures on the part of the robot \cite{tolmeijer_2020}.

In the following sections, we will summarize findings regarding trust violation and autonomy dissatisfaction along with their potential subsequent repair mechanisms. Derived from these findings, hypotheses will be formulated for examination in this work.  

\subsubsection{Trust Violation}

Robot performance is a key determinant of trust in HRI \cite{perkins_2022, zhang_2023}. Many constraints must be overcome to achieve optimal performance; however, disruptions due to error cases are unavoidable \cite{zhang_2023, sebo_2019}. Trust as a dynamic concept implies that it is constantly changing over time \cite{antonelli_2019, esterwood_2021, esterwood2022literature, esterwood_2023}. While human trust in cobots is crucial for successful HRI, it is often compromised by trust violations \cite{desai_2013, esterwood2022literature, robinette_2015}. Such violations can arise due to inevitable errors committed by cobots \cite{esterwood_having_2022, esterwood2022literature, robinette_2015, pompe_2022}. These violations are events that can negatively influence the perception of trustors regarding the trustworthiness of trustees \cite{esterwood_having_2022, esterwood2022literature, perkins_2022}. Research shows, the extent to which trust is lost is influenced by various factors, such as personal relevance, timing, the type of trust violation and the severity of the error \cite{kraus_2023, rossi}. Notably, the loss of trust appears to have a detrimental effect on cooperation between humans and cobots, underscoring the significance of this issue.  \cite{alarcon_2020}. 

In the field of trust repair studies, three distinct types of trust violations have been identified: capability violations, integrity violations and benevolence violations. These categories have been discussed in various academic publications \cite{rogers_2023, zhang_2023, esterwood2022literature}. In the case of capability-related violations of trust, the robot fails to meet the expectations that a human has of the robot's performance, as outlined in the literature \cite{esterwood2022literature, perkins_2022, pompe_2022, rogers_2023, sebo_2019}. When integrity is violated, a robot fails to meet the expectations that a human has of the robot's honesty and ethical consistency \cite{esterwood2022literature, sebo_2019}. Trust violations regarding benevolence occur when a robot fails to fulfil the expectations a human has about its purpose, thereby conveying a sense of malevolence \cite{esterwood2022literature}. Each of the three types of trust violation listed has been demonstrated to result in a reduction in trust on the part of the trust giver \cite{esterwood2022literature}. Regardless of the specific nature of the trust violation, an error of a cobot impacts trust. The dynamic transition from initial to knowledge-based trust underscores the temporal nature of trust development and its vulnerability to disruptions caused by trust violations. Therefore, we formulate the following hypotheses:\\\\
\noindent
\textbf{H1: }Knowledge-based trust after interacting with a faulty cobot is lower than initial trust before the interaction.\\
\textbf{H2: }A cobot's failure negatively impacts trust in the cobot.\\ 
\textbf{H3: }The more severe a cobot's failure is, the higher the negative impact of failure on trust in a cobot. 

\subsubsection{Autonomy Dissatisfaction} 
Friedmann hypothesized that a system's capability, complexity, fluidity and representation could promote or hinder a user's autonomy \cite{friedman1996value}. When a technological system fails, its capabilities are inevitably reduced for the time being. Consequently, system failure has the potential to impair one's need for autonomy by reducing perceived control after a failure \cite{guo2016role}. Furthermore, system failure might also reduce one's sense of choice by eliminating potential ways to achieve the desired outcome, forcing the user to pursue a specific path (like emergency protocols) or, in an extreme case, preventing one's goal from being fulfilled. Last, system failure could increase the mental load or perceived stress to a point where the user feels incapable of processing new or relevant information, resulting in a loss of (perceived) control \cite{bitkina2021user}. This effect would likely be more pronounced in severe failure situations. Additionally, a severe failure, especially by a cobot could pose real or perceived physical risks to the human user forcing them to adopt reactive rather than proactive roles to avoid harm. The following hypotheses are thus formulated on the basis of the argumentation presented:\\\\
\noindent
\textbf{H4: }A cobot's failure thwarts the need for autonomy.\\
\textbf{H5: }The more severe a cobot's failure is, the lower the satisfaction of a user's need for autonomy. 

\subsubsection{Trust and Autonomy Repair Through Transparent Communication}
In the event of failures occurring, the question of how to mitigate the negative effects on trust and autonomy arises.

One approach to dealing with trust violations entailed by cobot failure is to deploy trust repair strategies \cite{esterwood2022literature}. In order to design trustworthy robots, it is necessary to equip them with interaction strategies that increase transparency and predictability and enable an informed assessment of the extent of trustworthy and reliable task performance \cite{kraus_2023}. 

Transparency plays a crucial role in building trust in HRI \cite{akalin_2022, hancock_2011, kessler_2017, rubagotti_2022, schneewind1998invention}. It encompasses the level of information provided to users in order to make the process and the actions of the robots understandable and to achieve appropriate use of the system \cite{kessler_2017}. Transparent communication strategies are particularly important after errors have occurred \cite{kraus_2023}. However, the implementation of such strategies is case-specific and requires an individualised approach \cite{schoett_2023}. A previous study has shown that it is not only the content of an explanation that is decisive but also other aspects of the conversation, such as the language style or formality \cite{schoett_2023}. A fundamental recommendation for robot design is to ensure users are informed about the performance, reliability, functionality and capability of robots, while taking their emotional needs into account \cite{hancock_2011}. It is imperative that any information presented is read and internalized by the recipient in order to facilitate effective trust repair. \cite{nayyar_2018}.

To restore one's satisfaction with autonomy after a robot or cobot failure, transparent communication might also play an important role. Employee empowerment, a concept closely linked to workers' autonomy satisfaction was found to be positively influenced by transparent communication from peers and supervisors by providing useful job-related information. This enables the workers to actively participate in the decision-making process \cite{lee2023transparent}. 

Within SDT, the sub-theory of cognitive evaluation explores the impact of behaviour-initiating or regulating external events on competence and autonomy. External events can encompass various forms such as rewards (verbal or material), feedback or evaluations. Ryan and Deci propose that events that convey a feeling of self-determined competence promote an intrinsic locus of causality, thereby satisfying the need for autonomy \cite{ryan2018}. We argue, that simple transparent communication after a cobot's failure can convey a feeling of self-determined competence by enabling the users to decide on the next possible steps to take, enhancing their sense of control. Additionally, providing a rationale for specific system actions might promote understanding and endorsement even in contexts that lie beyond the user's direct control \cite{peters2023wellbeing}.   

In accordance with the aforementioned theoretical considerations, the findings of Yu et al. are consistent with the hypothesis that visualising the predicted path of a human can result in a loss of sense of agency, potentially due to the perception of being controlled by a robot \cite{yu2023your}. Given the considerations regarding transparent communication as a way to restore trust and autonomy satisfaction the following hypotheses are proposed:\\\\
\noindent
\textbf{H6: }Transparent communication after a cobot's failure restores trust in the cobot.\\
\textbf{H7: }Transparent communication after a cobot's failure restores satisfaction of the need for autonomy.

\section{Method}\label{Method}
The present study set out to investigate the impact of failure severity on trust perception and autonomy satisfaction, and to assess transparent communication as a potential means of recovering these perceptions after a failure. To this end, a VR experiment was conducted in a 2x2 factorial design. The following sections provide a detailed description of the study design, the methodology, which includes the VR environment, the study measures, the procedure, as well as the data preparation, analysis and sample description.

\subsection{Experimental Study Design}
\begin{figure}[H]
    \centering
        \includegraphics[scale=.25]{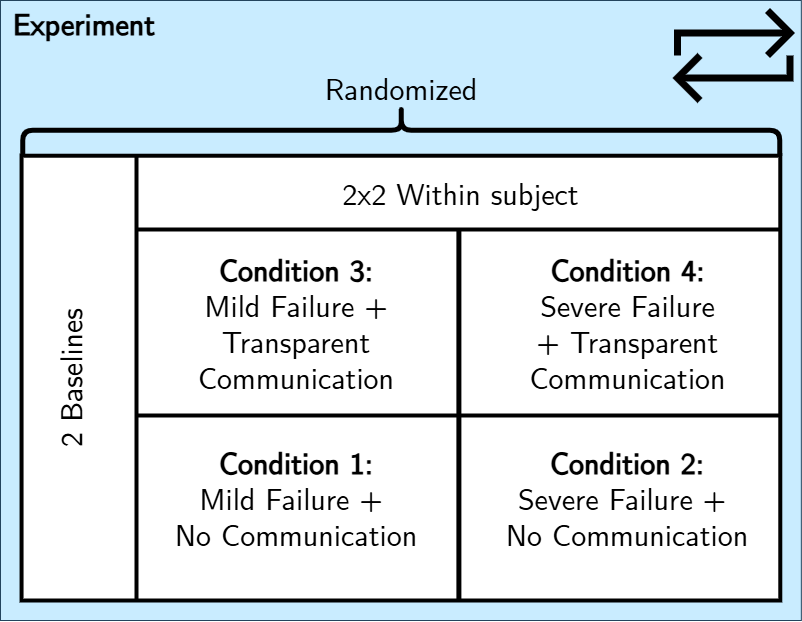}
    \caption{The 2x2 factorial experiment design used to test the effect of failure severity and transparent communication on trust and autonomy satisfaction}
    \label{FlowChart}
\end{figure}

The study design consists of a 2x2 experimental design. The incorporation of two baseline conditions as a control group and four treatment conditions as a treatment group ensures the maintenance of the experimental manipulation within a within-subject design. The control and treatment groups allow us to examine the statistically relevant distinction between the failure-free and manipulated conditions. The randomisation of the order of the baseline and treatment conditions serves to balance fatigue and order effects (see figure \ref{FlowChart}). Concerning randomization, both the experimenter and the test subjects were blinded. 

The subjects' task was to retrieve a package of specific content from a shelf jointly with the cobot and place it on a work table by signalling to the cobot to retrieve the package. 

\begin{figure}[H]
    \centering
        \includegraphics[scale=.14]{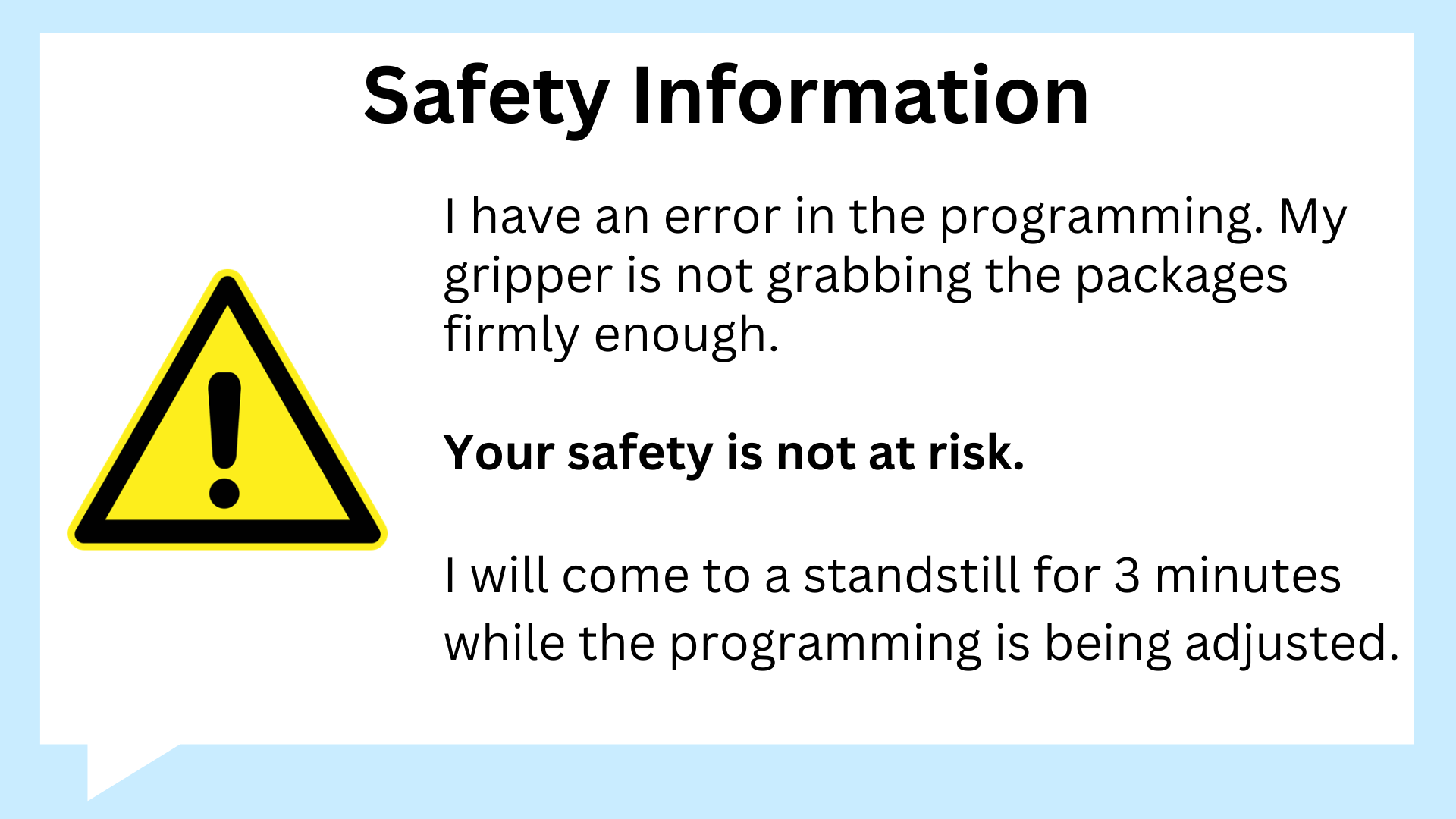}
    \caption{Information displayed on a screen beside the cobot in condition 3 after the failed attempt of the cobot to grab the package}
    \label{Small_Failure}
\end{figure}

The experimental manipulation involved the presentation of two distinct cases, as displayed in figure \ref{FlowChart}: one pertaining to a low \textit{failure severity} and the other to a high failure severity (mild vs. severe failure). Additionally, the presentation of information following a failure was also manipulated (information presentation vs. no information presentation), operationalizing \textit{transparent communication}. In the mild failure case, the non-functioning gripper of the cobot was imagined as the result of incorrect programming, which caused the cobot to cease functioning. The condition of minor failure and transparent communication was accompanied by information regarding safety on a screen, as displayed in figure \ref{Small_Failure}. 

\begin{figure}[H]
    \centering
        \includegraphics[scale=.14]{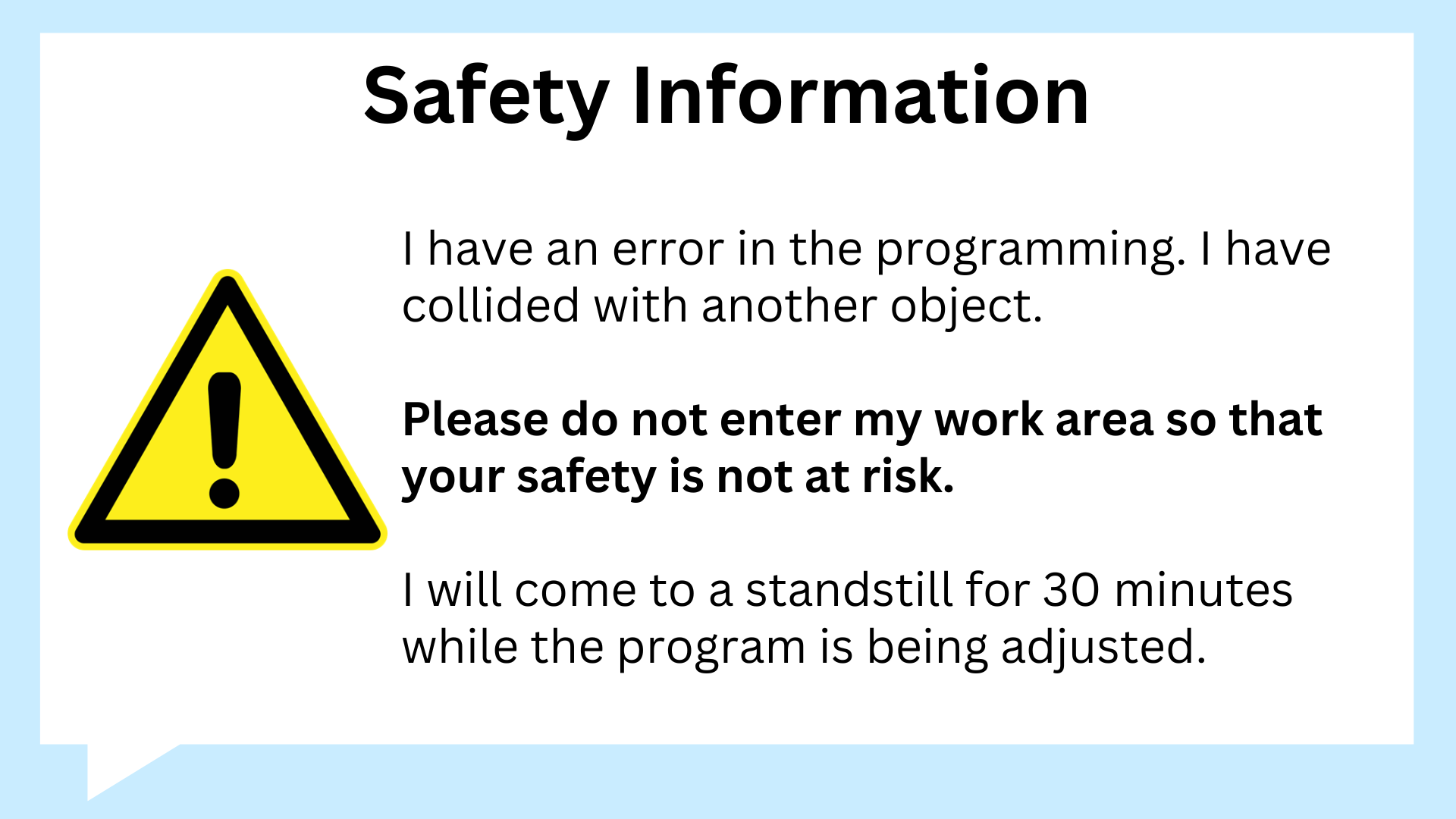}
    \caption{Information displayed on a screen beside the cobot in condition 4 after the cobot knocked over the shelf}
    \label{Serious_Failure}
\end{figure}

In the severe failure case, a collision occurred between the cobot and another shelf in the warehouse, resulting in the latter's subsequent fall. The failure was imagined as the result of a programming error, which caused the cobot to cease functioning. The severe failure and communication condition included slightly different information according to the cobot's failure (see figure \ref{Serious_Failure}). 

In all cases, the test subjects were required to complete the task with or without assistance from the cobot.
After each condition, the dependent variables of \textit{trust} and \textit{autonomy} were measured.

\subsection{Virtual Reality Environment}
\begin{figure}[H]
    \centering
        \includegraphics[scale=.18]{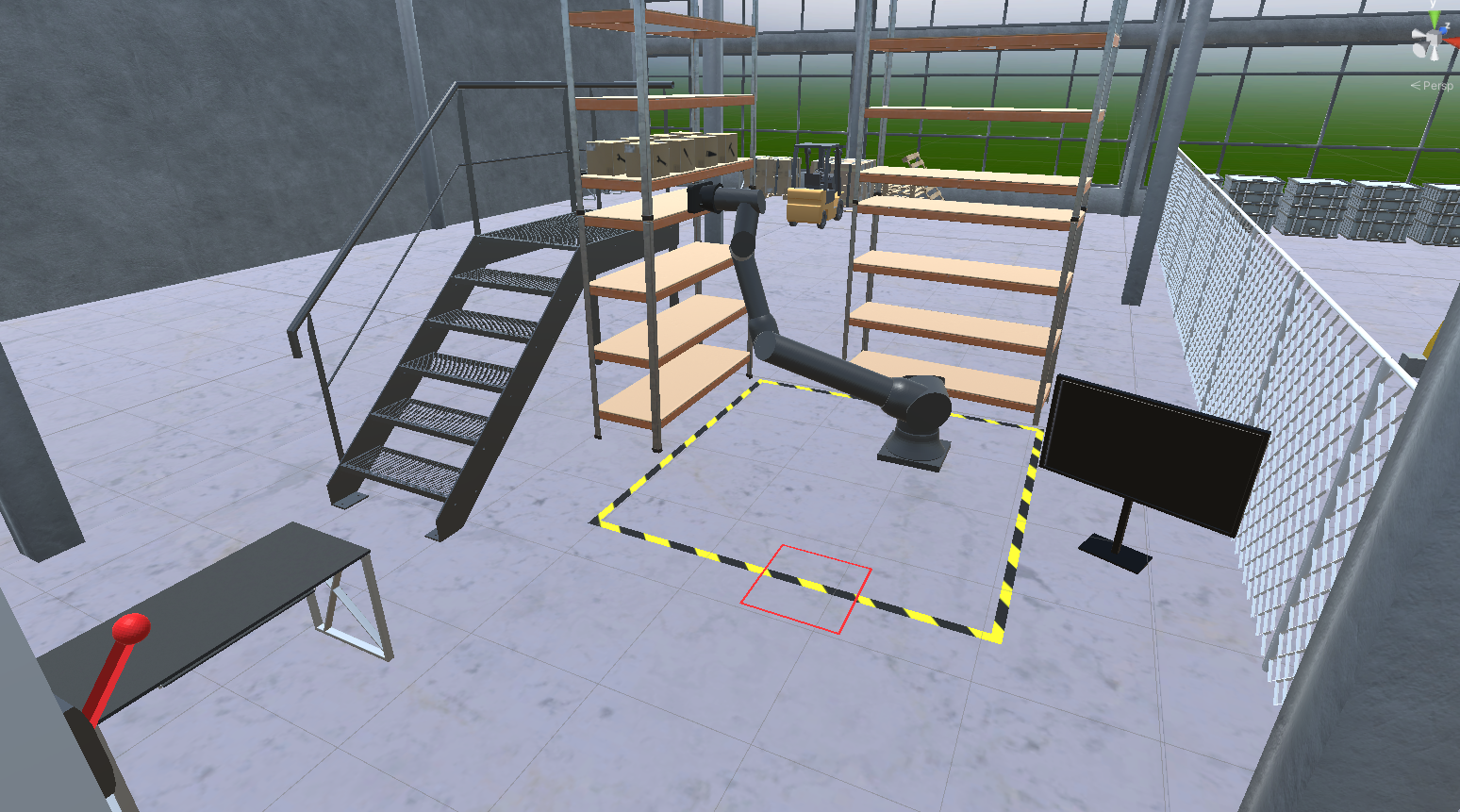}
    \caption{The virtual warehouse in which the subject performed the pick and place task is displayed on the left, with the red lever (which activates the cobot) and the desk (on which the package had to be placed) also visible. The cobot is situated in its operating area in the middle of the scene, with a staircase leading to the shelf containing the packages to be picked.In conditions 3 and 4 of severe failure, the cobot knocks over the shelf on the right. The right side of the scene displays the screen, which is utilised to present pertinent information in the event of a failure in conditions 3 and 4}
    \label{exp_image_1}
\end{figure}

The study design was conducted within a highly controlled, yet as realistic as possible, VR environment. To this end, a warehouse was programmed and built in Unity (Engine) Version 2022.3.4f1 to serve as an industrial workplace (see figure \ref{exp_image_1}). 

The utilisation of VR technology facilitates the creation of immersive and controlled environments, thereby enabling precise manipulation of independent variables and facilitating realistic interaction \cite{hald_2021}. To ensure full immersion in the experimental environment, subjects used a VR headset (HTC Vive Pro) and controllers for navigation and interaction with the virtual environment, thereby enhancing the ecological validity of the study. 

The VR workspace comprised a starting point and a screen that served to set the task of the study. The cobot was situated centrally within a delineated area of the work environment, comprising three axes and an end effector for package gripping. Its appearance resembled that of an industrial cobot, standing at a height of approximately a few metres. A secondary screen was positioned near the cobot, functioning as its interface. 
Two shelves were present, one of which was accessible via a staircase and the other was used for collisions in the event of severe failures. A work table was also located in the work area, and the cobot could be given a signal to collaborate via a red lever located on a column in the work area. 

\begin{figure}[H]
    \centering
        \includegraphics[scale=.13]{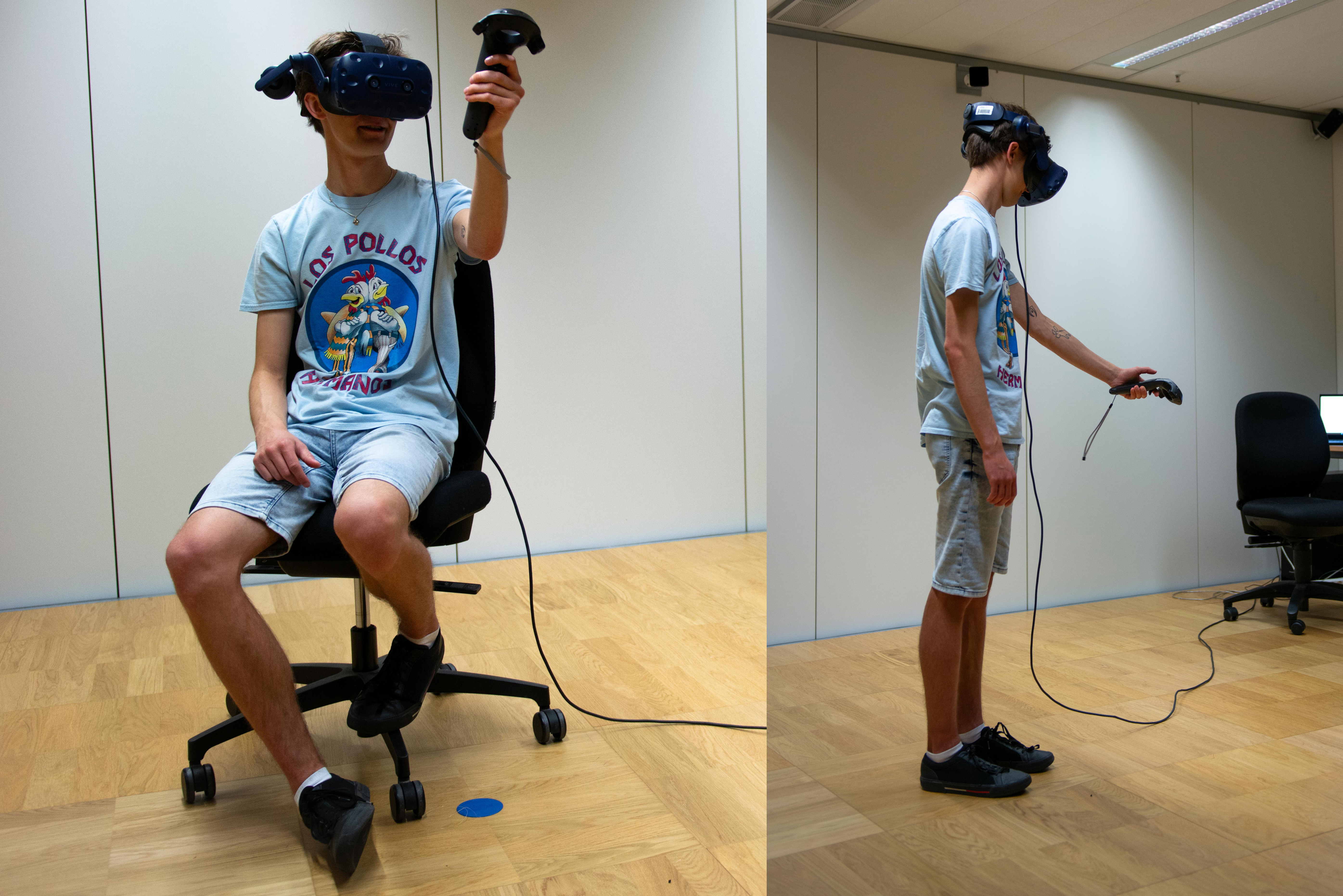}
    \caption{Subject within the laboratory setting in a seated and standing position}
    \label{exp_image_2}
\end{figure}

Test subjects could use the controller in their hands to move within the range of movement defined by sensors either sitting or standing (see figure \ref{exp_image_2}). The subject could choose their position based on their preference to reduce cybersickness symptoms. 

\subsection{Study Measures}
In the present study, the baseline and the experimental conditions (failure severity and transparent communication) were varied as within-factors, with the evaluation of the tasks functioning as the dependent variables (trust and autonomy). We queried the subject`s demographic data and explanatory user factors. Multiple-choice questions were used to collect the demographic information. We measured the psychometric scales using six-point Likert-scale items ranging from $1 = $ ‘no agreement at all’ to $6 = $ ‘full agreement’. To enhance the reliability and validity of the data collected, we developed the questionnaire based on existing and validated instruments. McDonald's omega ($\omega$) was calculated to test the internal reliability of the scales. 

\textbf{\textit{Demographic data}} of the subjects was assessed by means of a series of questions concerning the subjects' age, gender, highest level of education, experience of working in production, experience with cobots, and the experience was assessed to check for confounding effects.\\
We measured the \textbf{\textit{dependent variable}} trust using a, - to our use case adapted (e.g. \textit{The collaborative robot is trustworthy.}), - six-item Likert scale on trust in dealing with automated systems (TIAS) \cite{pohler2016itemanalyse}. For consistency, we surveyed the full 11-item scale, which also measures the dimension of distrust, but the analysis was conducted exclusively on the trust dimension. The need for autonomy scale developed by Van den Broeck et al. consisting of six items was adapted (e.g. \textit{I feel like I was able to be myself in this task.}) and used to assess the need for autonomy \cite{van2010capturing}. \\ 
The measured \textbf{\textit{exploratory user factors}} included affinity for technology interaction (ATI) \cite{franke2019personal} and autonomy orientation derived from the causality orientation scale by Deci and Ryan (1985) \cite{deci1985general}. Furthermore, the manifestation of cybersickness symptoms post-interaction was measured using the scale developed by Kourtesis et al. \cite{kourtesis2023cybersickness} we measured as well. However, due to the suboptimal reliability of the autonomy orientation scale, it was decided to exclude autonomy orientation as a factor without further analysis. 

\subsection{Study Procedure}
\begin{figure}[H]
    \centering
        \includegraphics[scale=.2]{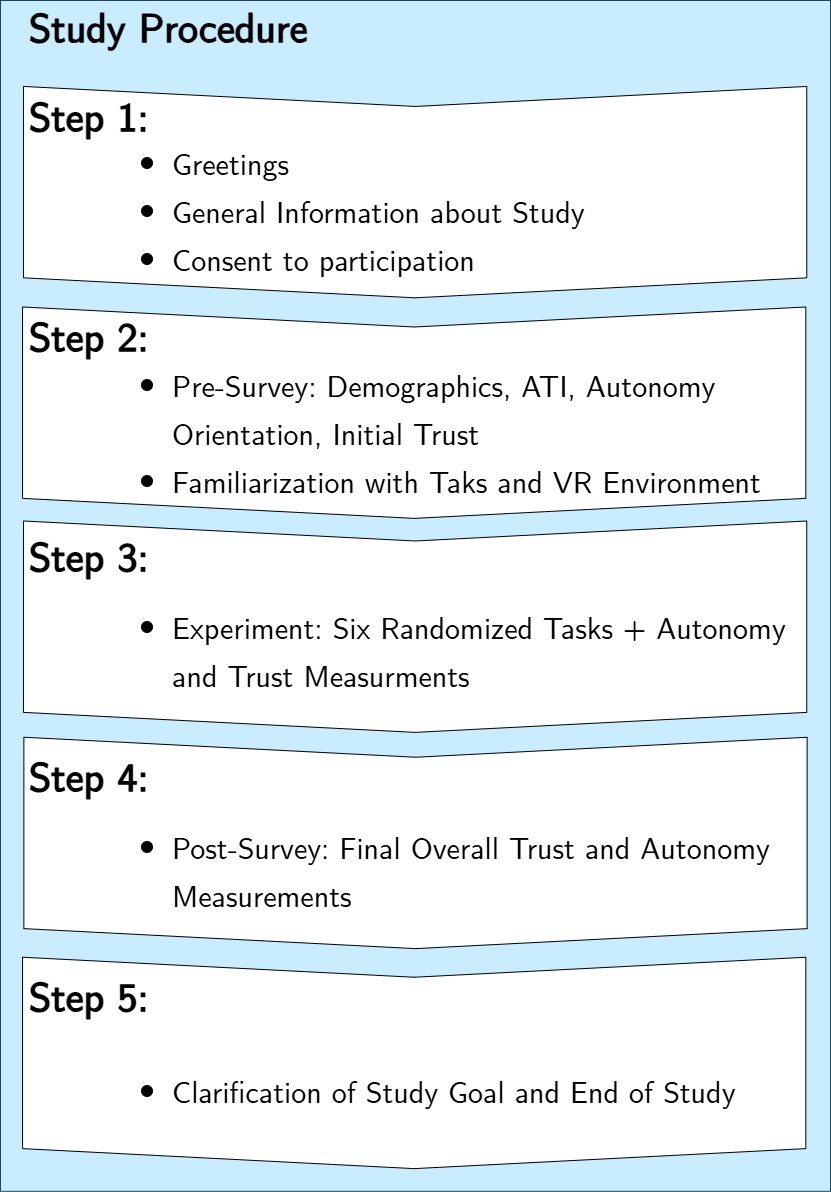}
    \caption{Study procedure displayed in five steps}
    \label{FlowChart1}
\end{figure}

The experimental study was conducted at the \blackout{Human-Computer Interaction Center of RWTH Aachen University in Germany}. The experimenter and the test subject were both present during the study. The experiment was performed in five stages, as illustrated in figure \ref{FlowChart1}.

1) The subjects were greeted and escorted to the laboratory where the experiment was to be conducted. Participants were provided with an information sheet detailing the objective, nature, and procedure of the study, along with any potential risks and data protection measures. We omitted information regarding occurrences of failure and the recovery of trust and autonomy to minimise any potential bias regarding their effects. The potential risks of pressure marks caused by the VR headset and symptoms of cybersickness (headache, nausea and vertigo) were communicated to the participants, who then provided written consent for participation. 

2) After receiving the written consent, participants completed a preliminary survey encompassing their demographic data, ATI and autonomy orientation scale as well as their initial trust towards cobots. Throughout the experiment, subjects responded to the questionnaire on a tablet. For the initial trust assessment, subjects were presented with a printed image of the cobot employed in the experiment. 

Following this, the experimenter provided an introduction to the experiment, explaining the task i.e. picking the right box with the cobot's aid, and again, the procedure was explained. Subsequent to this, the experimenter then introduced the participant to the VR area. Participants were then familiarised with the VR environment and controls. They were given the choice to either stand or sit during the experiment. They were given the option of either standing or sitting during the experiment to reduce any potential cybersickness symptomes. Participants were then given the opportunity to practice moving within the VR environment before undertaking two non-failure trial tasks. In the initial trial, participants were instructed to utilise the cobot, while in the subsequent trial, they were prompted to use the staircase as an alternative route. It was emphasised that, under all circumstances, the cobot should be used in the forthcoming experimental conditions. Furthermore, participants were informed that they were at liberty to pause the experiment at any time, particularly in the event of experiencing symptoms of cybersickness. To ensure the well-being of participants and to mitigate the occurrence of cybersickness symptoms, water and sweets were made available. 

3) After the trial tasks and a potential break, participants started with the main experimental block. the participants proceeded to the primary experimental phase. A total of six tasks were administered, with two tasks designated as baseline and four tasks assigned to the treatment conditions. Participants were requested to respond to the trust and autonomy scales between tasks and were encouraged to take a break, drink, and eat. 

4) After all six tasks and their respective trust and autonomy measurements, participants evaluated their final overall trust and rated their perceived cybersickness symptoms. 

5) Finally, the participants were provided with written clarification regarding the objective of the study, namely the examination of the impact of failures on trust and the need for autonomy. They were encouraged to ask any further questions and were then thanked for their participation.

\textbf{Task Description}: The objective of each task, irrespective of the control or treatment condition, was to retrieve the correct package from the shelf and place it on the workstation. The package to pick was stated on the display located at the start point. Packages were distinctively marked by tool symbols printed on their sides. To collaborate with the cobot, participants activated the red lever. If the task was a control condition, the cobot grasped the correct package and transferred it to the participant, who then placed it on the workstation. Conversely, in a treatment condition, participants were required to utilise the staircase to retrieve the package from the shelf autonomously. 

\subsection{Data Preparation and Analysis}\label{Data Preparation}

For data analysis and preparation we used Python 3.12.3 with packages pandas, numpy, scipy and statsmodels, IBM SPSS Statistics 29.0.0.0 to calculate effect sizes for the conducted ANOVAs and Jamovi Version 2.3.28.0 to calculate McDonald's omega. Data from Qualtrics, i.e. the subjects' responses, and Unity, containing information on the order of experimental conditions, were merged. Test cases and two instances where software problems occurred during testing were omitted from the dataset, resulting in a $n = 39$ dataset. Items with statements contradicting the content of the construct were re-coded to match the direction of the other items. Some items had a shifted scale due to a labelling error in Qualtrics and were recoded to match the intended scale. The internal consistency of the measured constructs was examined using McDonald's omega ($\omega$) which is comparable to Cronbach's alpha ($\alpha$) but is based on more realistic assumptions \cite{Hayes02012020}.  All constructs, except autonomy orientation, showed satisfying values of $\omega \geq 0.70$ (see table \ref{constructs}). A mean score was calculated over all items of one construct and for every respondent. For the trust and autonomy measures after the baselines a mean value was computed over both measurements after the respective baselines. Due to the scale's nature of measuring perceived cybersickness symptoms, the respective scale's consistency did not need to be examined. Responses for the cybersickness scale were added to an overall score (see section \ref{sample_description}). The data were analyzed using descriptive and inference statistical methods with a significance level set at 5\%. To determine the appropriate sample size, a power analysis was conducted using G*Power Version 3.1.9.7. For a medium effect of $f = 0.25$, the significance level at 5\% and a power of 0.95 the required sample size was estimated to be 36 \cite{cohen2013statistical}. 
Although trust and the need for autonomy are found to be related \cite{heyns2018volitional}, we opted to perform separate ANOVAs for both dependent variables instead of one MANOVA to emphasize the univariate focus of this work \cite{huang2020manova}. 

\begin{table}[ht]
\caption{Descriptive data for all constructs with their respective mean ($M$), standard deviation ($SD$), and McDonald's omega ($\omega$)}\label{constructs}%
\sisetup{
	detect-all,
    table-format = +.2
}
\begin{tabular}{l@{} c@{} S[table-format=+0.2]@{} S[table-format=+0.2]@{} S[table-format=+0.2]@{}}
\toprule
Construct & Source & {$\omega$} & {M} & {SD}\\
\midrule
Technology affinity & \cite{franke2019personal} & 0.89 &4.39&0.71  \\\hline
Initial trust & \multirow{8}{*}{\cite{pohler2016itemanalyse}} & 0.85 &3.91 & 0.79 \\
Knowledge-based trust & & 0.88 &2.96 & 0.98 \\
Trust baseline 1 & & 0.92 &4.21&1.06  \\
Trust baseline 2 & & 0.92 &3.87&1.25  \\
Trust condition 1 & & 0.92 &2.79&1.18  \\
Trust condition 2 & & 0.87 &2.35&0.91  \\
Trust condition 3 & & 0.89 &2.93&1.06  \\
Trust condition 4 & & 0.88 &2.76&1.06  \\\hline
Autonomy baseline 1 & \multirow{6}{*}{\cite{van2010capturing}} & 0.89 &4.50&1.02  \\
Autonomy baseline 2 & & 0.84 &4.18&1.05  \\
Autonomy condition 1 & & 0.79 &3.51&0.94  \\
Autonomy condition 2 & & 0.84 &3.34&1.07  \\
Autonomy condition 3 & & 0.76 &3.71&0.93  \\
Autonomy condition 4 & & 0.78 &3.64&0.98  \\
\botrule
\end{tabular}
\end{table} 

\subsection{Sample Description}\label{sample_description}
The final sample that was acquired featured $n = 39$ participants with a mean age of 25.62 years ($SD = 4.40$). The age range of the participants was from 20 to 40 years. The study was open to participation by individuals aged 18 years or over. 20 (51.2\%) of the participants identified as female, while the remainder identified as male. The overall education of the sample was high, with $24$ (61.5\%) posessing a university degree, and $15$ (38.4\%) a high school certificate.
The majority of the participants ($n = 36, 92.3\%$) did not currently work in production or had any prior experience with cobots ($n = 33, 84.6\%$). However, almost half of the sample ($n = 21, 53.8\%$) reported prior experience with VR. The sample's affinity for technology ($M = 4.39, SD = 0.71$) was rather high. The prevalence of cybersickness symptoms experienced after the interaction was low to medium ($M  = 15.46, SD = 6.36$), with a minimum possible value of six and a maximum possible value of 38. No correlation was found for age and gender, indicating a balanced sample concerning these variables.  

\section{Results}\label{Results}
In the subsequent section, the results of the hypothesis stated in section \ref{Related Work} will be presented. Prior to analysis, the data was checked for correlations between the dependent variable measurements of autonomy and trust, and user variables such as experience with robots, cybersickness, technical affinity, experience with cobots, experience with VR, and wether they are working in production. No correlations were found between those variables and the dependent variables. The dependent variables of trust and the need for autonomy satisfaction exhibited weak to moderate correlations, with three significant results (see Table \ref{Correlation_DVs}).

\begin{table}[ht]
\caption{Spearman correlations between the measurements of trust and need for autonomy satisfaction after the experimental conditions (n = 39)}\label{Correlation_DVs}%
\sisetup{
	detect-all,
    table-format = +.2
}
\begin{tabular}{l@{} S[table-format=+0.2]@{} S[table-format=+0.2]@{} S[table-format=+0.2]@{} S[table-format=+0.2]@{} S[table-format=+0.2]@{}}
\toprule
\textbf{Constructs} & {Trust B.  } & {Trust 1  } & {Trust 2  } & {Trust 3  } & {Trust 4  }\\
Autonomy B. & 0.58** & 0.31 & 0.39 & 0.41 & 0.39\\
Autonomy 1 & 0.22 & 0.41 &0.37 & 0.24 & 0.29  \\
Autonomy 2 & 0.15 & 0.35 & 0.40 & 0.20 & 0.26  \\
Autonomy 3 & 0.18 & 0.23 & 0.21 & 0.34 & 0.29   \\
Autonomy 4 & 0.29 & 0.51* & 0.16 & 0.42 & 0.53* \\
\botrule
\end{tabular}
\footnotetext{\footnotesize{*** significant at $p < .001$, ** significant at $p < .01$, * significant at $p < .05$ (Bonferroni corrected); B.: baseline; 1: Mild Failure + No Information; 2: Severe Failure + No Information; 3: Mild Failure + Information; 4: Severe Failure + Information}}
\end{table} 

\subsection{Impact of Failures on Trust and Autonomy}
\begin{figure}[H]
    \centering
        \includegraphics[scale=.55]{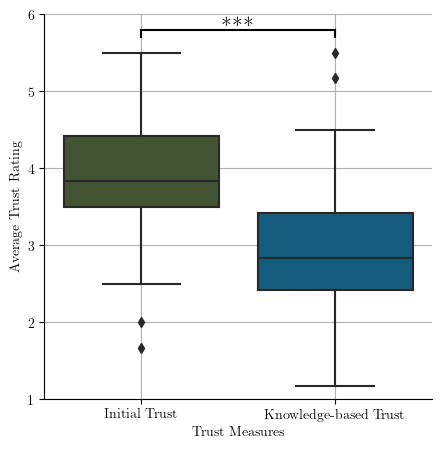}
    \caption{Boxplots comparing the measures of initial and knowledge-based trust illustrating the distribution, median, interquartile range, and outliers. Brackets and stars indicate the significant difference between initial trust and knowledge-based trust (n = 39)}
    \label{fig:trust_ik_boxplot}
\end{figure}

Initial trust was rated medium to high ($M = 3.91, SD = 0.79$) and knowledge-based trust medium to low ($M = 2.96, SD = 0.98$) (see figure \ref{fig:trust_ik_boxplot}). A paired sample t-test was conducted to test if the trust in the cobot is lower after the interaction with the faulty cobot than before an interaction happened. Results of the test yield a significant decrease of trust after the interaction compared to trust before the interaction with a medium effect size ($t(38) = 6.52, p < .001, d = 1.04$). Therefore, hypothesis one \enquote{Knowledge-based trust after interacting with a faulty cobot is lower than initial trust before the interaction.} is confirmed.

\begin{figure}[H]
    \centering
        \includegraphics[scale=.5]{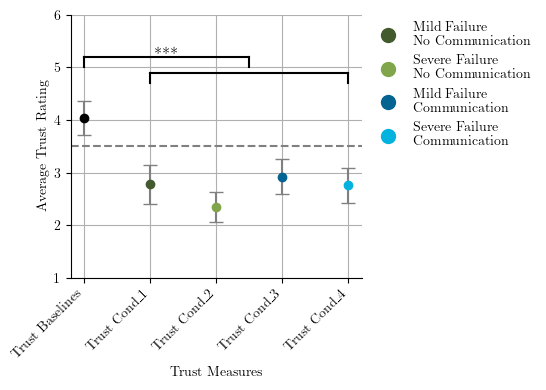}
    \caption{Comparison of mean trust measures for the baseline and failure conditions including confidence intervals. The dotted line represents the scale midpoint. Brackets and stars indicate the significant difference between trust measured after the baselines and trust measured after the failures (n = 39)}
    \label{fig:trust_boxplot}
\end{figure}

A mean value was computed over both trust measures after the baseline conditions ($M = 4.04, SD = 1.02$) and a mean value for the trust measures after the failure conditions ($M = 2.71, SD = 0.92$) (see figure \ref{fig:trust_boxplot}). A paired sample t-test was conducted to test if the trust in the cobot suffers after a failure occurs compared to after non-faulty interactions. Results of the test yield a significant decrease of trust for the failure conditions compared to the baseline conditions with a great effect size ($t(38) = 9.54, p < .001, d = 1.53$). Therefore, hypothesis two \enquote{A cobot's failure negatively impacts trust in the cobot.} is confirmed.

\begin{figure}[H]
    \centering
        \includegraphics[scale=.5]{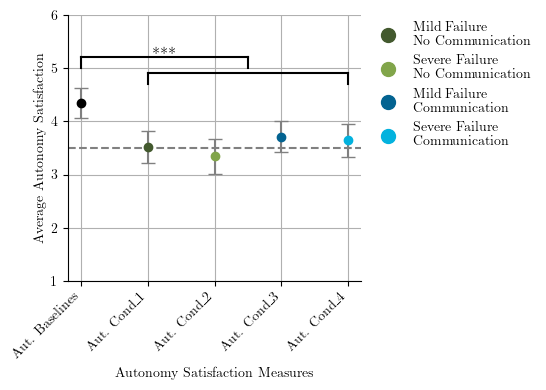}
    \caption{Comparison of mean autonomy measures for the baseline and failure conditions including confidence intervals. The dotted line represents the scale midpoint. Brackets and stars indicate the significant difference between trust measured after the baselines and trust measured after the failures (n = 39)}
    \label{fig:autonomy_boxplot}
\end{figure}

A mean value was computed over both autonomy satisfaction measures after the baseline conditions ($M = 4.34, SD = 0.93$) and a mean value for all autonomy satisfaction measures after the failure conditions ($M = 3.55, SD = 0.83$) (see figure \ref{fig:autonomy_boxplot}). A paired sample t-test was conducted to test if autonomy satisfaction suffers after a failure occurs compared to after non-faulty interactions. Results of the test yield a significant decrease in autonomy satisfaction for the failure conditions compared to the baseline conditions with a medium effect size ($t(38) = 5.57, p < .001, d = 0.89$). Therefore, hypothesis four \enquote{A cobot's failure thwarts the need for autonomy.} is confirmed.

\subsection{Failure Severity and Transparent Communication}

To assess the influence of failure severity and transparent communication on trust a two-way repeated measures ANOVA was conducted. Assumption of normality was tested using the Shapiro-Wilk test and was violated for condition two and three. However, with a sample size of 39 and the other trust measurements being normally distributed, we assume trust measurements for conditions two and three to feature a normal distribution as well. The assumption of sphericity was given. Results revealed a significant, large main effect of failure severity ($F(38) = 9.44, p < .01, \eta_p^2  = 0.199$) as well as transparent communication ($F(38) = 6.36, p < .05, \eta_p^2= 0.143$) on trust into the cobot. There is no interaction effect between failure severity and transparent communication ($F(38) = 1.42, p = 0.24, \eta_p^2= 0.036$). Figure \ref{fig:trust_boxplot} illustrates a negative influence of failure severity and a positive influence of transparent communication on trust. Therefore, both hypotheses three and six can be confirmed. Therefore, hypothesis three \enquote{The more severe a cobot's failure is, the higher the negative impact of failure on trust in a cobot.} as well as hypothesis six \enquote{Transparent communication after a cobot's failure restores trust in the cobot.} is confirmed.\\\\ 

To assess the influence of failure severity and transparent communication on autonomy need satisfaction a two-way repeated measures ANOVA was conducted. The assumption of normality and sphericity was found to be satisfied. Results revealed no effect of failure severity ($F(38) = 1.38, p = .25, \eta_p^2= 0.035$) on autonomy need satisfaction. A large significant effect was observed for transparent communication ($F(38) = 6.46, p < 0.05, \eta_p^2= 0.145$) and no interaction effect between failure severity and transparent communication ($F(38) = 0.26, p = 0.61, \eta_p^2= 0.007$). Figure \ref{fig:autonomy_boxplot} illustrates a positive influence of transparent communication on autonomy need satisfaction. Therefore, hypothesis five \enquote{The more severe a cobot's failure is, the lower the satisfaction of a user's need for autonomy.} is rejected and hypothesis seven \enquote{Transparent communication after a cobot's failure restores satisfaction of the need for autonomy.} can be confirmed.

\section{Discussion}\label{Discussion}
In this study, we investigated how cobot failures in HRC affect human perception and how the cobot's behaviour after a failure can restore trust and autonomy to allow for future optimal collaboration and workers' well-being. To this end, an experiment was conducted within a VR environment, to elucidate the relationship between cobot design, interaction, human perception and well-being. 

\noindent
\textbf{Trust.}
In the context of trust, the findings indicate that failures in cobots can lead to a reduction in trust. Subjects' knowledge-based trust following an interaction with a faulty cobot is significantly lower than their initial trust prior to the interaction. These findings suggest that trust may be subject to a dynamic nature that is susceptible to evolution through interaction with cobots as trustees \cite{kopp_2024, tolmeijer_2020, kok_2020, pompe_2022, lewis_2018}. Furthermore, the findings corroborate the notion that failures in HRC exert a detrimental influence on human trust, a phenomenon that has been documented in numerous studies, including those conducted by Desai et al. or Robinette et al. \cite{desai_2013, robinette_2015}. This outcome can be explained by the impairment of the trust-influencing factors of system performance, predictability and reliability \cite{nayyar_2018, tolmeijer_2020, baker_2018, hancock_2011, malle_2021}. 

Our study adds to the extensive body of research by highlighting that failure severity amplifies the negative effect of cobot failures on trust, similar to the findings of Rossi et al. and Aliasghari et al. \cite{rossi, aliasghari2021effect}. Although the outcome for participants was the same in both severity situations, the more severe failure might lead to the perception of a greater impairment of system performance. Furthermore, the occurrence of a major malfunction in the present study, namely the collision of the cobot with another object, has the potential to result in an elevated sense of concern regarding physical integrity.

Additionally, we found that transparent communication following a violation of trust exerts a positive influence on trust, in accordance with the results of Kraus et al. \cite{kraus_2023}. was unable to fully restore trust, which may indicate the presence of additional factors or other aspects of the communication that require further adaptation. For instance, Schött et al. have demonstrated that the language style and formality employed can influence the perception of information \cite{schoett_2023}. 

At this juncture, it is also imperative to emphasise the necessity for adequate levels of trust. As Nayyar and Wagner assert, there is a propensity to trust in automated systems such as cobots, even in the absence of prior experience, which can engender the risk of overtrust \cite{nayyar_2018}. The findings suggest that moderate initial trust is present but diminishes considerably throughout the interaction.

\noindent
\textbf{Autonomy.}
While autonomy satisfaction suffers less than trust after a failure, people still experience a thwarting of their psychological need for autonomy. As Friedman hypothesised, this may be due to a reduced perceived capability of the cobot that in turn hinders the user from achieving their desired goals \cite{friedman1996value}. However, in this study it was still possible to achieve the goals after a failure, leading us to conclude that the perception of loss of control or choice may have thwarted autonomy. The requested solution to the task, i.e. the package that the cobot handed to the user, was not possible after the failure, forcing the user to take the only remaining solution. It could be argued that by requesting a specific solution in advance, there was no choice in the first case. However, a study of Belgian prisoners showed that the perception of choice, which is not necessarily the same as real choice, is positively associated with autonomy satisfaction \cite{van2017choosing}. This reinforces the idea that the mere perception of choice is more important than the actual choice itself.  

Our finding did not support the hypothesis that failure severity negatively influences the user's need for autonomy. This can be explained by the impact of choice perception on autonomy. In both failure cases, the outcome is the same for the user in terms of the subsequent actions required. Therefore, the perceived choice may not be further hindered and, consequently, the satisfaction of the need for autonomy is not further reduced.  

However, transparent communication after a failure could partially restore autonomy satisfaction in both cases. This may be because transparent communication acts as a form of autonomy-supportive behaviour, providing the user with a rationale for the failure and the subsequent steps, leading to feelings of feelings of acknowledgement and a reduced sense of control over the situation \cite{moreau2012importance, peters2023wellbeing}. Like in the case of trust, transparent communication alone did not suffice to restore the thwarted autonomy satisfaction after a failure fully. This may be because, although participants were informed of the next steps, they were no longer in charge and were compelled to resolve the task in the manner dictated by the circumstances, which was probably not in line with their original intentions.

\subsection{Limitations and Future Research}
Despite the novel insights into the effects of failures and transparent communication on trust and autonomy, this work faces some limitations, which will be discussed in this section. Based on those limitations and the previous discussion we provide recommendations for future research on this topic. 

Firstly, it is important to note that the sample is representative of a gender-balanced, yet young and highly educated population. Therefore, it is not possible to apply the results to the general population. However, the low variance of the sample highlights the strong effect of the experimental manipulations. It is noteworthy that only three participants were currently employed in manufacturing, the primary domain where cobots are employed \cite{schmidbauer2023empirical}. This observation suggests that the assigned task may not align with the subjects' daily responsibilities, potentially failing to reflect their interests and values. Consequently, the failures and automation by the cobot may not lead to the anticipated conflict with the subjects' need for autonomy, which in turn could lead to an even greater loss of autonomy. Regarding the impact of transparent communication on autonomy, Peters suggests that user diversity might play a role in the amount of information and explanation needed to stand as a meaningful rationale for the cobot's behaviour \cite{peters2023wellbeing}. Future research could further examine the influence of if and how resolving the occurring failures by themselves might help to maintain the need for autonomy, or whether the information provided needs to be more specifically tailored to the target audience. We therefore encourage replication of this study with a larger sample and with a sample that resembles the target population. 

This study builds on the idea that knowledge-based trust is developed after the human-cobot interaction. However, it is crucial to recognise that the development of trust in a real-world setting is a more protracted process than in the controlled environment of an experiment. A longitudinal study could offer a more comprehensive analysis of knowledge-based trust and assess the reliability of knowledge-based trust as posited by McKnight \cite{mcknight_2011}. Nayyar and Wagner explain that users must read information truthfully to internalize it \cite{nayyar_2018}. The experimental design does not preclude the possibility that test subjects did not perceive the transparent communication immediately or did not read it conscientiously, and therefore there is a possibility that the effect of the transparent communication might be even greater than found. Consequently, subsequent studies could ascertain whether communication is perceived and understood.

We found no relationship between our control variables like cybersickness or experience with VR and the dependent variables. The flexibility and controllability of the VR environment were advantageous; nevertheless, the effect that physical interaction might have on the perception of the cobot should not be overlooked \cite{li2019comparing, bainbridge2008effect}. It is therefore recommended that experiments similar to the present study be conducted with real cobots to confirm the results.

\section{Conclusion}\label{Conclusion}
Cobots are already a significant technological advancement that is driving the industrial evolution towards Industry 4.0. Their inherent flexibility, cost-effectiveness and seamless integration into the workflow of small companies are anticipated to drive their widespread adoption in the coming years \cite{kildal2018potential, marketshare}. While Industry 4.0 is still in its early deployment phase, the human-centred approach of Industry 5.0 and the idea of designing technology for human well-being have received considerable attention \cite{barata2023industry, peters2018designing, peters2023wellbeing}. This study contributes to human-centred work environments by examining the impact of system failures and transparent communication on human trust in cobots and autonomy need satisfaction. These factors are instrumental in fostering a production that functions as a safe, healthy and accepted workplace. It can be concluded that while minor or severe failures are inevitable in the context of technology and human-robot interaction, the subsequent communication and information provided play a pivotal role in shaping the user's perception of the technology, their workplace, and their overall well-being. 

\backmatter

\bmhead{Data availability}

The full dataset, Jupyter Notebook, SPSS and Jamovi Syntax used for analysis, as well as the complete VR project and questionnaire are publicly accessible at the Open Science Framework repository \url{https://osf.io/we6vj/?view_only=eee3f8cd5a7e48cb9d46022661e17e8e}. 

\bmhead{Funding}

Funded by the Deutsche Forschungsgemeinschaft (DFG, German Research Foundation) under Germany’s Excellence Strategy – EXC-2023 Internet of Production – 390621612.

\section*{Declarations}

\bmhead{Conflict of interests}
The authors have no competing interests to declare that are
relevant to the content of this article.

\bmhead{Informed consent}
Informed consent was obtained from all participants
in the study.

\bibliography{08_references}

\begin{sidewaystable*}[ht]
\caption{Overview of items used to measure the constructs including the items' mean ($M$) and standard deviation ($SD$). (R): Items recoded for analysis. }\label{SingleItem}%
\sisetup{
	detect-all,
    table-format = +.5
}
\begin{tabular}{l@{ } l l l l l }
\toprule
Measured onstructs and their respective single items & \makecell[l]{Baseline:\\ M \textit{(SD)}} & \makecell[l]{Cond.1:\\ M \textit{(SD)}} & \makecell[l]{Cond2:\\ M \textit{(SD)}} & \makecell[l]{Cond.3:\\ M \textit{(SD)}} & \makecell[l]{Cond.4:\\ M \textit{(SD)}}\\
\midrule
Trust in Cobot (after Baselines and failure conditions)& & & & & \\
\midrule
The collaborative robot provides safety. & 4.81 \textit{(1.20)}& 3.23 \textit{(1.35)}& 2.59 \textit{(1.25)}& 3.36 \textit{(1.35)}& 3.03 \textit{(1.33)}\\
The collaborative robot works flawlessly. & 4.63 \textit{(1.30)}& 1.90 \textit{(1.37)}& 1.74 \textit{(1.27)}& 1.87 \textit{(1.22)}& 1.85 \textit{(1.29)}\\
The collaborative robot is reliable. & 4.83 \textit{(1.16)}& 2.33 \textit{(1.40)}& 1.95 \textit{(0.97)}& 2.49 \textit{(1.39)}& 2.49 \textit{(1.52)}\\
The collaborative robot is trustworthy. & 3.58 \textit{(1.57)}& 3.08 \textit{(1.48)}& 2.46 \textit{(1.27)}& 3.38 \textit{(1.46)}& 3.05 \textit{(1.32)}\\
I can trust the collaborative robot. & 4.83 \textit{(1.26)}& 3.03 \textit{(1.46)}& 2.54 \textit{(1.25)}& 3.33 \textit{(1.42)}& 3.18 \textit{(1.39)}\\
I am familiar with the collaborative robot. & 4.28 \textit{(1.25)}& 3.15 \textit{(1.53)}& 2.85 \textit{(1.35)}& 3.13 \textit{(1.40)}& 3.00 \textit{(1.41)}\\
\midrule
Need for Autonomy Satisfaction (after Baselines and failure conditions)& & & & &\\
\midrule
I feel like I was myself at this task. & 4.41 \textit{(1.32)}& 4.10 \textit{(1.35)} & 3.77 \textit{(1.31)}& 4.46 \textit{(1.10)}& 3.85 \textit{(1.44)}\\
In this task, I felt as if I had to follow the commands of others. (R)& 4.10 \textit{(1.43)}& 3.82 \textit{(1.39)}& 3.85 \textit{(1.37)}& 3.82 \textit{(1.41)}& 3.82 \textit{(1.43)}\\
If I could choose, I would have done things differently in this task. (R) & 4.73 \textit{(1.30)}& 3.51 \textit{(1.52)}& 3.28 \textit{(1.67)}& 3.90 \textit{(1.54)}& 3.85 \textit{(1.58)}\\
The tasks I had to do were in line with what I really wanted to do. & 3.62 \textit{(1.61)}& 2.62 \textit{(1.27)}& 2.38 \textit{(1.27)}& 2.67 \textit{(1.30)}& 2.56 \textit{(1.29)}\\
I felt free to do the task the way I thought it could best be done. & 4.29 \textit{(1.41)}& 3.10 \textit{(1.35)}& 2.87 \textit{(1.47)}& 3.36 \textit{(1.53)}& 3.51 \textit{(1.54)}\\
In this task, I felt forced to do things I did not want to do. (R) & 4.90 \textit{(1.24)}& 3.92 \textit{(1.48)}& 3.90 \textit{(1.70)}& 4.08 \textit{(1.55)}& 4.26 \textit{(1.37)}\\
\midrule
Affinity for Technology & {M \textit{(SD)}} & & & &\\
\midrule
I like to occupy myself in greater detail with technical systems.  & 4.56 \textit{(1.05)} & & & &\\
I like testing the functions of new technical systems. & 4.97 \textit{(0.81)}& & & &\\
I predominantly deal with technical systems because I have to. & 4.51  \textit{(1.17)}& & & &\\
When I have a new technical system in front of me, I try it out intensively. & 4.64 \textit{(0.93)}& & & &\\
I enjoy spending time becoming acquainted with a new technical system. & 4.41 \textit{(1.09)}& & & &\\
It is enough for me that a technical system works; I don’t care how or why. & 4.08 \textit{(0.98)}& & & &\\
I try to understand how a technical system exactly works. & 4.08 \textit{(0.90)}& & & & \\
It is enough for me to know the basic functions of a technical system. & 3.74 \textit{(1.16)}& & & &\\
I try to make full use of the capabilities of a technical system. & 4.51 \textit{(0.88)}& & & &\\
\midrule
Initial \& Knowledge-Based (K.B.) Trust & \makecell[l]{Initial:\\ M \textit{(SD)}} & \makecell[l]{K.B.:\\ M \textit{(SD)}} & & &\\
\midrule
The collaborative robot provides safety. & 4.13 \textit{(1.00)}& 3.05 \textit{(1.32)} & & &\\
The collaborative robot works flawlessly. & 4.05 \textit{(1.00)}& 2.10 \textit{(1.12)}& & &\\
The collaborative robot is reliable. & 4.51 \textit{(1.12)}& 2.72 \textit{(1.23)} & & &\\
The collaborative robot is trustworthy. & 4.23 \textit{(0.99)}& 3.21 \textit{(1.28)} & & &\\
I can trust the collaborative robot. & 4.23 \textit{(1.06)}& 3.18 \textit{(1.30)} & & &\\
I am familiar with the collaborative robot. & 2.28 \textit{(1.23)}& 3.49 \textit{1(.25)} & & &\\
\botrule
\end{tabular}
\end{sidewaystable*}

\end{document}